\documentclass{iopjournal}

\usepackage{amsmath,amssymb,mathrsfs}
\usepackage{amsbsy} 
\usepackage{graphicx} 
\usepackage{float}

\usepackage[colorlinks=true, allcolors=blue]{hyperref}

\def\eqref#1{(\ref{#1})}
\def\be{\begin{equation}}
\def\ee{\end{equation}}
\def\bpm{\begin{pmatrix}}
\def\epm{\end{pmatrix}}
\def\bea{\begin{eqnarray}}
\def\eea{\end{eqnarray}}
\def\dd#1#2{_{#1#2}}
\def\uu#1#2{^{#1#2}}
\def\ud#1#2{{^{#1}}_{#2}}
\def\du#1#2{{_{#1}}^{#2}}
\def\half{\tfrac 12}
\DeclareMathOperator{\bnabla}{{\pmb{\nabla}}}
\def\grad{\bnabla}
\def\rot{\bnabla\times\,}
\def\divergence{\bnabla\cdot\,}
\def\vac#1{\mathbf #1}

\long\def\textout#1{}

\begin{document}

\articletype{Paper} 

\title{On some nonlinear extensions to Maxwell's electrodynamics}

\author{Bertrand Berche$^{1}$\orcid{0000-0002-4254-807X} and Ernesto Medina $^{2}$\orcid{ }}

\affil{$^1$Laboratoire de Physique et Chimie Th\'eoriques, Universit\'e de Lorraine -- CNRS, Nancy, France}

\affil{$^2$Departamento de F\'isica, Colegio de Ciencias e Ingeniería, Universidad San Francisco de Quito, Diego de Robles y Via Interoceánica, Quito 17901, Ecuador}

\email{bertrand.berche@univ-lorraine.fr}


\begin{abstract}
Soon after Maxwell's celebrated synthesis of electrodynamics, the discovery of the electron raised fundamental questions about the nature and structure of elementary particles. The picture of a small, rigid, charged body of finite radius faced the difficulty of introducing non-electromagnetic forces or stresses to prevent the charge from dispersing, an assumption that was increasingly regarded as unsatisfactory. The alternative description in terms of a point charge was no more satisfactory, because its electromagnetic self-energy diverges. From the early twentieth century onward, several authors therefore explored modifications of Maxwell's theory, with nonlinear electrodynamics emerging as a particularly natural and radical possibility. In this paper, we revisit at the undergraduate level some of the nonlinear theories proposed in this context and explore their ability to address the classical self-energy problem of charged particles. 
\end{abstract}

\section{Introduction}

Maxwell's theory of electrodynamics is a masterly construction that, by the end of the 19th century, synthesized electricity, magnetism, and electromagnetic wave phenomena, including light. Maxwell's Treatise\cite{maxwell1873treatise} is one of the most influential texts in the history of physics. Its accessible form and formal structure --- with a clear description of the basic experiments and empirical postulates, together with a presentation of the mathematical objects and methods suited to their deployment --- also most certainly contributed to its pedagogical impact. Indeed, most present-day textbooks remain imbued with this presentation, discussing experiments, electrostatics, magnetostatics, induction, polarized and magnetized media, and a synthesis with the Maxwell equations and the theory of light. Modern texts often modernize the form rather than the substance.

Despite the authority commanded by the edifice of Maxwell's theory, the end of the 19th century and the beginning of the 20th saw the emergence of conceptual difficulties that called into question its applicability at very small length scales. The self-energy problem is emblematic of this. Indeed, the electron was then assumed to be a point-like charged particle producing a field varying as $1/r^2$, the Coulomb field, to which is associated an energy density decreasing as $1/r^4$, whose integral over all space diverges at the lower bound (the point charge limit) and yields an infinite total energy.

 The hypothesis that the particle has a very small but finite size was put forward at the turn of the century by Thomson\cite{Thomson1881}, studying the electromagnetic contribution to the mass: ``{\em The effect of the electrification is the same as if the mass of the sphere were
increased by $4 m e^2/(15a)$}''. This resolves the divergence difficulty affecting the self-energy, but raises further questions when considered later within the framework of special relativity: a rigid body is incompatible with relativity, and the cohesion of the object requires the introduction of forces of non-electromagnetic origin, the Poincaré stresses, prompting Landau and Lifshitz\cite{Landau1980Classical} to remark: ``{\em We can conclude that electrodynamics as a logically closed physical theory presents internal contradictions when we go to sufficiently small distances.}.''

At the beginning of the 20th century, theories competing with Maxwell electrodynamics were proposed in order to resolve the divergence of the Coulomb field at short distances and thereby accommodate the notion of a point charge. As we shall see, however, the earliest of these proposed extensions raise significant difficulties of their own. We restrict ourselves to the framework of electrostatics to examine several of these competing theories. We analyze other types of extensions of Maxwell's theory of electrodynamics in Ref.~\cite{Berche2025-10}.

The target audience consists of upper-undergraduate students who already have a basic understanding of tensor formalism and variational calculus. The paper begins with
a self-contained introduction to Maxwell electrodynamics in tensor notation and presents the problem of self-energy at an elementary level using classical electrostatics in spherical symmetry, before explaining the historical and pedagogical value of the nonlinear theories of Mie and Born–Infeld.

\section{The structure of Maxwell's theory}\label{sec2}
In this section, we give a very short overview of Maxwell's theory of electrodynamics. We use the tensorial approach and assume here that spacetime is described by a Minkowski manifold with metric tensor $\eta\dd\alpha\beta$ allowing the definition of the interval squared
\be ds^2=\eta\dd\alpha\beta dx^\alpha dx^\beta\equiv dx_\beta dx^\beta.
\label{Eq-ds2}\ee
In this notation, $\alpha,\beta$ are Lorentz indices, taking values $0$ for time and $1,2,3$ for space variables in Cartesian coordinates.
Repeated in their covariant and contravariant positions, the Lorentz indices are summed over all values $0,\dots 3$, and an expression like \eqref{Eq-ds2} is called a {\em contraction}.
 The infinitesimal $dx^\alpha$ is a contravariant $4$-vector,  $(dx^\alpha)=(c dt,dx,dy,dz)^\top$ where $\top$ means transposed, i.e. a column vector. We choose the signature $(\eta\dd\alpha\beta)=\mathrm{diag}\,(+1,-1,-1,-1)$ so that the covariant row $4$-vector $(dx_\beta)={(cdt, -dx,-dy, -dz)}$. Under Lorentz transformations, $4$-vectors and tensors transform linearly, and contractions over all indices yield invariant quantities called Lorentz scalars (see Ref.~\cite{Franklin2010,Berche2025-4} for an undergraduate-level presentation).

The theory of Maxwell is a field theory based on the definition of a field-strength tensor\cite{Zangwill2013,Tong2025Electromagnetism,Berche2025-5}
\be F\dd\alpha\beta(x)=\partial_\alpha A_\beta(x) - \partial_\beta A_\alpha(x)\ee
in terms of a $4$-covector $A_\alpha(x)$ called the $4$-potential. The dependence on spacetime is denoted by $(x)$, a compact notation for the four spacetime coordinates, and {\em does not mean} $x^1=x$. In terms of ordinary scalar $\phi(\vac r,t)$ and vector $\vac A(\vac r,t)$ potentials, the $4$-potential reads, in contravariant form, as
\be
(A^\alpha)=\bpm \phi/c \\ \vac A\epm.
\ee
In terms of ordinary electric and magnetic fields
\be 
\vac E=-\grad\phi-\partial_t\vac A,\quad \vac B=\rot\vac A,
\ee 
the field-strength tensor has components 
\be 
(F\dd\alpha\beta)=\bpm 0 & E_x/c & E_y/c & E_z/c \\
-E_x/c & 0 & -B_z & B_y \\
-E_y/c & B_z & 0 & -B_x \\
-E_z/c & -B_y & B_x & 0 \epm
\ee 
{The commutativity of partial derivatives implies the Bianchi identity} 
\be
\partial_\alpha F\dd\beta\gamma+ 
\partial_\beta F\dd\gamma\alpha
+
\partial_\gamma F\dd\alpha\beta=0,
\ee
 equivalent to
\be
\rot\vac E+\partial_t\vac B=0,\quad \divergence\vac B=0.\label{Eq_maxwellpremiercouple}
\ee
{Thus these two Maxwell equations follow identically from the definition of $F\dd\alpha\beta$}, so no further assumption is needed. The two Maxwell equations above are therefore fundamental properties of the 
electric and magnetic fields.

The two remaining Maxwell equations can be deduced from an action principle. For this purpose, a natural candidate for an action compatible with Special Relativity, i.e., built from a Lorentz scalar --- a quantity in which all Lorentz indices are contracted ---   is obviously $F\dd\alpha\beta F\uu\alpha\beta$. 
This is enough for the free electromagnetic field, but it still lacks coupling to matter.
The sources are described, as the continuity equation suggests,
\be \partial_t\rho + \divergence\vac J = 0, \ee 
by a $4$-vector $J^\alpha$ called the $
4$-current density such that the equation above reads in tensor form as
\be \partial_\alpha J^\alpha=0.\ee
In this relation, the $4$-derivative has components $ \partial_\alpha=\partial_{x^\alpha}$, $( \partial_\alpha)=(\tfrac 1c\partial_t,{\pmb\nabla})$ and therefore $(J^\alpha)=(\rho c,\vac J)^\top$. The simplest coupling between the electromagnetic field and the external sources $J^\alpha$ is linear; it requires a $4$-covector to form a contraction, and the only candidate available is the $4$-potential $A_\alpha$, so the simplest acceptable theory must include a term in $A_\alpha J^\alpha$.
We can thus postulate an action --- known under the name of Schwarzschild action\cite{Schwarzschild1903a,Schwarzschild1903b} --- 
\be
S_{\rm EM}[A]=-\int d^4x \Bigl( \tfrac{1}{4\mu_0} F\dd\alpha\beta F\uu\alpha\beta + A_\alpha J^\alpha\Bigr) = \int d^4 x\, {\mathscr L}_{\rm EM}(A,\partial A)
\ee
where the coefficients are fitted to recover Maxwell's equations and the signs imposed by the signature choice.
The  missing piece of Maxwell's theory now follows from the variational principle
\be
\frac{\delta S}{\delta A_\beta}=
\frac{\partial  {\mathscr L}_{\rm EM}}{\partial A_\beta}-\partial_\alpha\Bigl(  \frac{\partial  {\mathscr L}_{\rm EM}}{\partial (\partial_\alpha A_\beta)}\Bigr)=
-J^\beta+\tfrac{1}{\mu_0}\partial_\alpha F\uu\alpha\beta=0
\ee 
and in terms of the ordinary electromagnetic fields, this equation corresponds to 
\be 
\divergence\vac E=\frac{\rho}{\varepsilon_0},\quad \rot\vac B-\tfrac{1}{c^2} \partial_t\vac E =\mu_0\vac J
\ee
where  $\varepsilon_0\mu_0 c^2=1$. 

Maxwell's theory also inherits an interesting property, gauge invariance, considered at the time as a mathematical curiosity. All four Maxwell's equations are indeed unchanged if we perform the following transformation of the $4$-potential:
\be
A_\alpha(x)\to A'_\alpha(x)=A_\alpha(x)+\partial_\alpha\chi(x)
\ee
(for any non-singular function $\chi(x)$, see Ref.~\cite{10.1119/1.4955153} otherwise) because the field-strength tensor is gauge invariant, $F\dd\alpha\beta(x)\to F'\dd\alpha\beta(x)=F\dd\alpha\beta(x)$ .
Translated into ordinary notation,   the combined transformations
\bea
&&\phi(\vac r,t)\to\phi'(\vac r,t)+\partial_t\chi(\vac r,t),\\
&&\vac A(\vac r,t)\to\vac A'(\vac r,t)-\grad\chi(\vac r,t)
\eea
keep $\vac E$ and $\vac B$ unchanged and therefore describe the same physics.

The self-energy problem arises when one considers the energy required to ``build'' a system of charges. This is a very well-known question treated in any textbook on electrodynamics\cite{Zangwill2013,griffiths_electrodynamics_2024,Tong2025Electromagnetism}, 
and the result is well known by students (see Supplementary Material (SM)):
\be
U_{\rm es}=\half \int_{\infty} \rho(\vac r) \phi(\vac r)\,d^3r=\half \varepsilon_0 \int_{\infty}
|\vac E(\vac r)|^2 d^3r.\label{Eq_ElectrostaticEnergyMaxwell}\ee 
The quantity $u_{\rm es}=\half \varepsilon_0
|\vac E(\vac r)|^2 $  is called the electrostatic energy density and $U_{\rm es}[\phi]$ is a functional of the potential $\phi$.
Let us nevertheless mention that this result  holds for Maxwell theory, but it is not generally valid and the form in terms of the electric field $\vac E$ relies on the use of the local Gauss theorem $\rho(\vac r)=\varepsilon_0\divergence\vac E(\vac r)$ in its standard form.

Now, consider the point charge. As we have discussed in the introduction, Maxwell's theory predicts a Coulombic electrostatic field associated with a point charge,
\be \vac E(\vac r)=\frac{q}{4\pi\varepsilon_0}\frac{\hat{\vac r}}{r^2}
\ee
where $r=|\vac r|$ and $\hat{\vac r}=\vac r/r$ is the unit vector pointing outwards from the charge. To this field corresponds an energy density
\be 
u_{\rm es}(\vac r)=\tfrac 12\varepsilon_0|\vac E(\vac r)|^2=\frac{q^2}{32\pi^2\varepsilon_0}\frac{1}{r^4},
\ee
which, integrated over the whole space, leads to the divergence at the lower bound,
\be
U_{\rm es}=\int d^3r\, u_{\rm es}(\vac r)=\frac{q^2}{8\pi\varepsilon_0}\Bigl[ \frac 1r \Bigr]^\infty_0 \to \infty.
\ee
This general structure of Maxwell's theory is the {\em simplest}, but consistent extensions that could resolve the divergence of the fields at the point charge have soon been proposed.

\section{The theory of Mie}\label{sec3}

\subsection{The strategy proposed by Mie}

Mie\cite{Mie:1912_Part1,Mie:1912_Part2,Mie:1913_Part3} proposed to repair the difficulty of the diverging self-energy within a unified field theory of matter, in which the electron itself would {\em emerge} as a stable particle-like solution of a nonlinear field theory, with no separate  ``point charge'' added by hand at all. 

As far as we know, besides Pauli's monograph\cite{Pauli:1981_Relativity} and Vizgin's book\cite{Vizgin1994}, there are not many expositions of nonlinear theories, but a notable exception is Ref.~\cite{BialynickiBirula1983} by Iwo Bia{\l}ynicki-Birula, where the situation regarding Mie's theory is clearly presented:
\begin{quotation}
``{\em In the classical models of the electron considered before Mie, the electron was not treated as a purely electromagnetic entity, but it was also ``made of other stuff'', like the Poincar\'e stresses and the mechanical mass. Mie wanted the electromagnetic field to be the only ingredient; everything else was to be derived from it. In particular, he wanted the electromagnetic current to come from electromagnetism. To achieve this goal, Mie assumed that the Lagrangian depends on the four-potential $A_\alpha$ not only through its four-dimensional curl $F\dd\alpha\beta$, but also on $A_\alpha$ itself. 
}''
\end{quotation} 

Let us expand the argument of Bia{\l}ynicki-Birula.
In ordinary Maxwell theory, the free-field Lagrangian depends only on $F_{\alpha\beta}$, so it is automatically gauge invariant, and charge must be added as an external source term $- A_\alpha J^\alpha$.
Mie's ambition was to have matter charge density, the electron itself, emerge automatically as a solution of the field equations, without adding any separate source term. 
But a nonzero, localized charge density cannot arise purely from the homogeneous field equations of a gauge-invariant Lagrangian built only from $F_{\alpha\beta}$. 
 Mie therefore proposed replacing Maxwell's Lagrangian, which is quadratic in the fields, with a more general nonlinear function, ${\mathscr L}_{\rm NL}$, of the field invariants built from combinations of the $4$-potential and the field strength tensor,
\be Q_1=A_\alpha A^\alpha,\quad Q_2=F_{\alpha\beta}F^{\alpha\beta},\quad Q_3= F\dd\alpha\gamma F\ud\gamma \beta A^\alpha A^\beta,\ \dots\ee 
This theory is posterior to the Special theory of Relativity and should therefore be consistent with Lorentz covariance, hence the need for an invariant Lagrangian density written as a fully contracted expression\footnote{See Ref. \cite{Pauli:1981_Relativity} for a discussion on the possible invariants, and in particular the note [20] which is important in the perspective of Born--Infeld's theory.}.
By nonlinear, we qualify the structure of the equation of motion in terms of the $4$-potential, which is the field degree of freedom, therefore, a theory like 
${\mathscr L}\sim -\tfrac 14 F^2-A^2$ would be linear (and known under the name of Proca~\cite{Proca:1936_MassiveField}).

Mie's strategy is as follows. We keep a term in $Q_2$ in the Lagrangian density --- the standard source-free Maxwell term in $-\tfrac 14 F^2$ --- and add supplementary $Q_1$, $Q_3$, \dots types of terms that will replace the interaction $-A_\alpha J^\alpha$. The external source $J^\alpha$  becomes useless, and with it the question of the {\em classical structure of the electron}, provided that the self-energy ceases to be divergent.
To be specific, consider the simplified case in which the Lagrangian density is a function of $Q_1$ and $Q_2$ only,
\be 
{\mathscr L}_{\rm NL}=-\tfrac{1}{4\mu_0}
F_{\alpha\beta}F^{\alpha\beta} - W(A_\alpha A^\alpha),
\label{Eq_LMiealphabeta}
\ee 
where  $W$ is still arbitrary. The ``kinetic'' term keeps its standard Maxwell form.
The demand for this ${\mathscr L}_{\rm NL}$ is obviously
that at large distances, i.e., in weak fields, the new theory should reduce to ordinary Maxwell theory, matching all known electromagnetic phenomena,
but at short distances, that is in strong fields, near where a point charge would sit, the role of nonlinearity should correct the divergence, so the total field energy  is possibly finite.

Mie 
allowed ${\mathscr L}_{\rm NL}$
 to depend on $A_\alpha$
 directly, precisely so that the modified field equations,
\be {2}\partial_\alpha\left(\frac{\partial {\mathscr L}_{\rm NL}}{\partial F_{\alpha\beta}}\right) = \frac{\partial {\mathscr L}_{\rm NL}}{\partial A_\beta} \equiv -s^\beta, \label{Eq_Lagrange}\ee
could generate, instead of an external $J^\beta$, an effective four-current $s^\beta$ out of the field itself. A ``matter core'' appears as a particular soliton-like solution where  $A_\alpha$ is large enough for the $Q_1$-,
($Q_3$- if they are present) -dependent terms to matter, surrounded by a region where $A_\alpha$ is small and the theory smoothly relaxes back to source-free Maxwell theory. 

\subsection{Expression of the electromagnetic energy}

Nevertheless, before considering the possible divergence of the energy, we must first discuss its precise form, since the definition of $u_{\rm es}$ in section~\ref{sec2} for Maxwell's theory no longer works in the absence of external sources. It is therefore useful to reformulate the argument using a field-theoretic approach.
There, the energy density follows from the definition of the stress-energy tensor $T\dd\alpha\beta$ and a proper definition of this quantity follows from the coupling of matter-energy with gravitation through Einstein's equations\cite{wald2022advanced,Berche2025-9}:
\be 
T\dd\alpha\beta=\frac{2}{\sqrt{-g}}\frac{\delta}{\delta g\uu\alpha\beta}(\sqrt{-g}{\mathscr L)}
=2\frac{\delta{\mathscr L}}{\delta g\uu\alpha\beta}-g\dd\alpha\beta{\mathscr L}.\label{Eq_Tab}
\ee 
The equation is written here in the general case of a spacetime with metric tensor $g\dd\alpha\beta$, but we return to the Minkowski metric $\eta\dd\alpha\beta$ for the specific applications treated later.
The energy density is given by the $00$ component of the stress-energy tensor, and the associated total energy follows from integration over the entire space:
\be 
u=T\uu 00,\quad U=\int_\infty T\uu 00 d^3r .
\ee

Applying the strategy for the theory of Mie, with the  Lagrangian density \eqref{Eq_LMiealphabeta},
we find the terms
\bea 
&&2\frac{\partial}{\partial g\uu\alpha\beta}(-\tfrac{1}{4\mu_0}F\dd\mu\nu F\uu\mu\nu)=-\tfrac{1}{\mu_0}F\dd\alpha\rho F\du\beta\rho,\\
&&2\frac{\partial W(A_\mu A^\mu)}{\partial g\uu\alpha\beta}=2W'(A^2)A_\alpha A_\beta,
\eea 
leading to the following expression:
\be 
T^{\rm NL}\dd\alpha\beta=-\tfrac{1}{\mu_0}F\dd\alpha\rho F\du\beta\rho+\tfrac{1}{4\mu_0}\eta\dd\alpha\beta F\dd\rho\sigma F\uu\rho\sigma -2W'(A^2)A_\alpha A_\beta+\eta\dd\alpha\beta W(A^2),
\ee 
where, for short, we denote $A^2=Q_1=A_\mu A^\mu=\phi^2/c^2-|\vac A|^2$, $W'(x)$ is the derivative wrt its argument $x$, and $\tfrac{1}{4\mu_0}F\dd\mu\nu F\uu\mu\nu=Q_2/(4\mu_0)
=\tfrac12\varepsilon_0c^2|\vac B|^2-\tfrac 12\varepsilon_0|\vac E|^2$.
The $00$ component follows,
$-\tfrac{1}{\mu_0}F\dd 0\rho F\du 0\rho=\tfrac{1}{\mu_0c^2}|\vac E|^2$, $\tfrac{1}{4\mu_0}\eta\dd 00 F\dd\rho\sigma F\uu\rho\sigma=\tfrac{1}{2}\varepsilon_0(c^2|\vac B|^2-|\vac E|^2)$ and it can be written
as
\be 
T^{\rm NL}\dd 00= 
\tfrac 12\varepsilon_0|\vac E|^2
+\tfrac12\varepsilon_0c^2|\vac B|^2
-2\frac{\phi^2}{c^2}W'(\phi^2/c^2-|\vac A|^2)+W(\phi^2/c^2-|\vac A|^2). \label{Eq_T00Gal}
\ee

\subsection{The electrostatic limit}

We wonder whether a nonlinear term built from $Q_1$ --- i.e., with a higher-than-quadratic power, or other nonlinear functions, in ${\mathscr L}$ --- can, at least partially, solve the problem of the diverging energy of the point charge. 
Let us consider a simple example of a possible theory satisfying Mie's criteria in the simple electrostatic case, where the $4$-potential reduces to its time component, the scalar potential in spherical coordinates
 \be (A_\alpha)=(\phi(r)/c,0,0,0),\ee
so that the invariants take the simplified expressions
\be  Q_1=\phi^2(r)/c^2, \quad 
{ -\frac{Q_2}{4\mu_0}}=\frac 12\varepsilon_0 \Bigl(\frac{d\phi(r)}{dr}\Bigr)^2.\ee   
Note that we keep the definition of the electrostatic field in terms of the potential
\be 
\vac E(\vac r)=-\grad\phi(\vac r),
\ee
since it follows from $\rot\vac E=0$, and does not depend on the form of the Lagrangian density.
Mie's theory modifies the Schwarzschild action but leaves the first two equations \eqref{Eq_maxwellpremiercouple} unchanged, thus, the expressions of the fields in terms of the potentials still hold.

The actual criterion that matters is to have a finite $\int d^3 r\,u_{\rm es}$.
We can therefore write
\be 
{\mathscr L}_{\rm NL}\Bigl(\phi,\frac{d\phi}{dr}\Bigr)=\frac 12\varepsilon_0 \Bigl(\frac{d\phi(r)}{dr}\Bigr)^2-V(\phi(r)),\label{eq-miegeneral}
\ee
with some function of the potential $V(\phi)=W(\phi^2/c^2)$. This is a genuinely nonlinear equation --- there is no source term inserted; any charge has to emerge from the solution itself.

More general situations could even be 
\be 
{\mathscr L}_{\rm NL}(\phi,\phi')=K(E^2(r))-V(\phi(r)),\quad\hbox{with}\quad
K(E^2(r))\to_{E\to 0}\half \varepsilon_0 E^2(r).\label{Eq-KEPE}
\ee 
For simplicity, we can call $K$ the ``kinetic'' energy and $V$ the potential.
This case modifies the form of the free-field Maxwell term and will be considered in Born--Infeld theory; therefore, we omit this in this section.

We must take care not to apply \eqref{Eq_Lagrange} blindly, because the coordinates are not Cartesian and we must account for the volume element. This being done,
varying wrt $\phi$ in Eq.~\eqref{eq-miegeneral} thus leads to
\bea
\frac{\delta}{\delta\phi}\int 4\pi r^2 dr\, \Bigl[
\frac 12\varepsilon_0 \Bigl(\frac{d\phi(r)}{dr}\Bigr)^2-V(\phi(r))
 \Bigr]&=&\varepsilon_0\frac{d^2\phi}{dr^2}+\frac {2\varepsilon_0}{r}\frac{d\phi}{dr}+V'(\phi)=0,
\eea
which is nothing but
 the generalized Poisson 
equation, here in spherical symmetry, and with $V'(\phi)=dV/d\phi$,
\be \nabla^2\phi(\vac r)=-\varepsilon_0^{-1}\frac{dV(\phi)}{d\phi}=-\frac{\rho_{\rm eff}(\vac r)}{\varepsilon_0}, \ee
or the associated
 Gauss law
\be \divergence\vac E=\frac{1}{\varepsilon_0}\frac{dV(\phi)}{d\phi}.\ee
 We can therefore define an effective charge density
\be \rho_{\rm eff}(\vac r)=\frac{d V(\phi)}{d\phi},\ee
which, according to Mie's program, is not an external source but is built directly into the field equations and the expected ``elementary particle'' has a charge $Q=\int d^3r \,V'(\phi)$. 
{Consequently, this imposes a constraint on the function \(V(\phi)\), which specifically requires that \(V'(\phi) > 0\) for \(\phi > 0\) and \(V'(\phi) < 0\) for \(\phi < 0\), meaning that the derivative shares the sign of the potential $\phi$ (which in turn matches the sign of the charge).}

We want to investigate the behavior of the total electrostatic energy.
In the electrostatic case, the 00 component of the stress-energy tensor \eqref{Eq_T00Gal} simplifies to
\be
T^{\rm NL}\dd 00= 
\tfrac 12\varepsilon_0|\vac E|^2-\phi V'(\phi)+V(\phi),
\ee
where $W(\phi^2/c^2)=V(\phi)$, and the total energy
\be 
U_{\rm NL}^{\rm es}[\phi]=\int_\infty d^3r\,\Bigl( \tfrac 12\varepsilon_0|\vac E|^2-\phi V'(\phi)+V(\phi)
\Bigr). \label{Eq-UphiVprime}
\ee 

If we now use the equation of motion in the form of the generalized Poisson equation $\varepsilon_0\nabla^2\phi=-\frac{dV}{d\phi}$, multiply by $\phi$ and integrate over space, we get\cite{pohozaev1965eigenfunctions}
\bea 
\int d^3r\, \phi\frac{dV(\phi)}{d\phi}=-\varepsilon_0\int d^3r\,\phi\nabla^2\phi&=&-{\varepsilon_0}\oint d\vac S\cdot(\phi\grad\phi)+\varepsilon_0\int d^3r\,|\grad\phi|^2\nonumber\\
&\to&\varepsilon_0\int_\infty d^3r\,|\grad\phi|^2,
\eea 
where the second line holds as the integration volume tends to infinity, provided that the surface term vanishes. It follows that 
\be
U_{\rm NL}^{\rm es}[\phi]=\int_\infty d^3r\, 
\Bigl[ V(\phi)  - \tfrac 12\varepsilon_0
|\vac E|^2
\Bigr].\label{Eq-Energy1}
\ee
Note that Maxwell's theory corresponds to {$V(\phi)=\rho\phi$} and, because of Eq.\eqref{Eq_ElectrostaticEnergyMaxwell}, we recover the usual formula.

Now we use 
Derrick's theorem\cite{Derrick:1964_NonlinearWave}
which applies to the {\em action}, therefore here to the static action
\be 
S_{\rm NL}^{\rm es}[\phi]
=
\int d^3r
\Bigl[ \tfrac 12\varepsilon_0
|\vac E|^2-
V(\phi) 
\Bigr].\label{Eq-StatS1}
\ee 
We assume stationarity $\frac{dS}{d\lambda}=0$ at $\lambda=1$ wrt to the scale transformation $r\to\lambda r$ with
\be 
S_{\rm NL}^{\rm es}[\phi(\lambda r)]=\lambda^{-1}\int d^3r
\tfrac 12\varepsilon_0
|\grad\phi|^2 -{\lambda^{-3}}\int d^3r\,
V(\phi), 
\ee 
and it follows   that 
\be
\int_\infty d^3r\, V(\phi)=\tfrac 13\int_\infty d^3r\, \tfrac 12\varepsilon_0|\nabla\phi|^2,
\ee
therefore
\be
U_{\rm NL}^{\rm es}[\phi]=-\tfrac{2}{3}\int d^3r\, \tfrac 12\varepsilon_0|\vac E(\vac r)|^2,\label{Eq-Energy3}
\ee 
while Maxwell's theory has a prefactor equal to $+1$, because in Eq.~\eqref{Eq-UphiVprime}, when $V(\phi)=\rho\phi$, $V(\phi)-\phi V'(\phi)$ simply cancels.
According to Mie, this energy defines the particle's mass $m=U_{\rm NL}^{\rm es}/c^2$, and the negative value is obviously problematic.

\subsection{Lane--Emden solutions}

Let us first contemplate the power-law case $V(\phi)=\varepsilon_0\lambda\phi^{2n}$ with $\lambda$ a constant. This satisfies the convexity criterion. In the family of equations of motion known as Lane--Emden equations, the case $n=3$ is of particular interest, because it accommodates singular, but also regular solutions at the origin\cite{Mach_2012} that are selected if we impose a boundary condition $\phi(r\to0)\to{\rm const}$.
Indeed, the nonlinear theory considered here is generated by a term cubic in
$Q_1$, or equivalently by a $\phi^6$ interaction in three spatial
dimensions; the corresponding radial equation belongs to the critical
$k-1=5$ Lane--Emden class.  This case is particularly interesting because,
in addition to the property of regularity at the origin, it admits an
exact closed-form \cite{Mach_2012}.

Let us study this case explicitly. 
We therefore consider  the nonlinear Lagrangian
\begin{equation}
 {\mathscr L}_{\rm NL}\Bigl(\phi,\frac{d\phi}{dr}\Bigr)
 =
 \frac{1}{2}\varepsilon_0
 \left(\frac{d\phi}{dr}\right)^2
 -
 \varepsilon_0\kappa\phi^6(r),
 \label{eq:mie_phi6_lagrangian}
\end{equation}
where $\kappa$ has dimensions of
$\mathrm{(potential)^{-4}(length)^{-2}}$, therefore it depends on a typical length scale and on a typical electric charge to be defined. Indeed, rather than introducing
an external point charge, we shall determine the normalization of
$\kappa$ by requiring that the regular field configuration carry a
prescribed total charge $Q_{\rm eff}$.

The Euler--Lagrange equation for a spherically symmetric configuration
is
\begin{equation}
 \frac{d^2\phi}{dr^2}+\frac{2}{r} \frac{d \phi}{dr}
 +6\kappa\phi^5=0.
 \label{eq:mie_phi6_eom}
\end{equation}
A power-law solution of Eq.~\eqref{eq:mie_phi6_eom} is obtained by
setting in the general solution (see SM) $p=\frac 12$ and $A=(24\kappa)^{-1/4}$
\begin{equation}
 \phi_{\rm sing}(r)=\frac{1}{(24\kappa)^{1/4}\sqrt r},
\end{equation}
and therefore diverges at the origin.  The corresponding electric
field,
\begin{equation}
 E_{\rm sing}(r)=-\frac{d\phi(r)}{dr}
 =
 \frac{1}{2\,(24\kappa)^{1/4}\,r^{3/2}},
\end{equation}
has a weaker singularity than the Coulomb field, $E_{\rm C}\sim r^{-2}$, but also diverges at the origin.

Substituting the singular solution of the $\phi^6$ Mie potential into  the stress-energy tensor, we obtain the remarkably simple expression for the electrostatic energy:
\bea  
&& U_{\rm sing}^{\rm es}
 ={
 4\pi}\int_0^\infty r^2u_{\rm NL}^{\rm es}(r)\,dr
 \propto
 -\int_0^\infty\frac{dr}{r},
\eea 
which is logarithmically divergent at both the origin and infinity.
The singular solution, therefore, does not provide a finite-energy
electrostatic configuration.

The route offered by the potential function $V(\phi)=\varepsilon_0\kappa\phi^6$ seems to be a dead end for solving the self-energy divergence problem. 
But, as we announced, the critical $k-1=5$ Lane--Emden equation also admits the exact regular
solution
\begin{equation}
 \phi_{\rm reg}(r)
 =
 \frac{B}{\sqrt{r^2+a^2}},
 \label{eq:mie_phi6_regular_ansatz}
\end{equation}
where $a$ is an arbitrary length scale.
This has the interesting Coulomb form at large distances, 
\be \phi_{\rm reg}(r)
\to_{r\to\infty}\frac Br\ee
i.e. $B=Q_{\rm eff}/(4\pi\varepsilon_0)$. This solves the Euler equation, since
one has
\bea
&&\nabla^2\phi=-\frac{3Ba^2}{(r^2+a^2)^{5/2}},\\
&&\varepsilon_0^{-1}V'(\phi)=\frac{6\kappa B^5}{(r^2+a^2)^{5/2}}
\eea
which yields $\kappa B^4=a^2/2$, or the useful combination $\kappa B^5=Q_{\rm eff}a^2/(8\pi\varepsilon_0)$  normalized in terms of the physical
total charge.
The corresponding electric field
\begin{equation}
 E_{\rm reg}(r)
 =
 \frac{Q_{\rm eff}}
 {4\pi\varepsilon_0}
 \frac{r}{(r^2+a^2)^{3/2}}.
 \label{eq:E_regular_Qeff}
\end{equation}
  is regular everywhere, Coulombic at large distances, and satisfies
$E_{\rm reg}(0)=0$. 
It reaches its maximum at
$  r_{\rm max}=\frac{a}{\sqrt{2}}$,
and for $r\gg a$ approaches the Coulomb field,
\begin{equation}
 E_{\rm reg}(r)
 =
 \frac{Q_{\rm eff}}
 {4\pi\varepsilon_0r^2}
 \left[
 1-\frac{3a^2}{2r^2}
 +O\left(\frac{a^4}{r^4}\right)
 \right].
\end{equation}
The effective charge density follows,
\begin{equation}
 \rho_{\rm eff}(r)
 =
 \frac{3Q_{\rm eff}a^2}
 {4\pi(r^2+a^2)^{5/2}},
 \label{eq:rhoeff_Qeff}
\end{equation}
it is positive, smooth, and finite at the
origin.
The associated total charge is
\begin{align}
 Q
 &=
 4\pi\int_0^\infty
 r^2\rho_{\rm eff}(r)\,dr
= 3Q_{\rm eff}a^2
\underbrace{ \int_0^\infty
 \frac{r^2\,dr}{(r^2+a^2)^{5/2}}}_{(3a^2)^{-1}}=Q_{\rm eff}.
\end{align}
Thus, the total electric charge is not introduced as an external
point source, but, as demanded, it emerges from the  field configuration.

We now explicitly calculate the total electrostatic self-energy.
The energy density obtained using any of the three expressions \eqref{Eq-UphiVprime}, \eqref{Eq-Energy1}, or \eqref{Eq-Energy3}, associated with Eq.~\eqref{eq:mie_phi6_lagrangian}
is
\begin{equation}
u_{\rm reg}(r)
    =
    \frac{Q^2}{32\pi^2\epsilon_0}
    \frac{r^2-5a^2}{(r^2+a^2)^3}
    .
 \label{eq:energy_density_regular}
\end{equation}
The total self-energy is consequently
\begin{align}
 U_{\rm reg}^{\rm es}
 &=
 4\pi\int_0^\infty
 r^2u_{\rm reg}^{\rm es}(r)\,dr
=-\frac{Q^2}
 {12\pi\varepsilon_0}
\underbrace{ \int_0^\infty
 \frac{r^4\,dr}{(r^2+a^2)^3}}_{3\pi/(16 a)}
  = -
 \frac{Q^2}
 {64\varepsilon_0a}
 .
 \label{eq:self_energy_integral}
\end{align}
Thus, the nonlinear $\phi^6$ theory removes the Coulomb self-energy
divergence and replaces it with a finite energy proportional to
$Q_{\rm eff}^2/(\varepsilon_0 a)$, {\em but a negative total electrostatic energy}!

\subsection{A logarithmic solution}

Another interesting solution entering  Mie's classification is given by the logarithmic potential $\phi_{\rm log}(r)=\phi_0\ln(1+a/r)$, which delivers an asymptotic Coulomb form at large distances, $\phi_{\rm log}(r)\sim_{r\to\infty} \phi_0a/r$, and justifies calling the constant $\phi_0=Q/(4\pi\varepsilon_0 a)$; hence
\be
\phi_{\rm log}(r)=\frac{Q}{4\pi\varepsilon_0 a}\ln\Bigl(1+\frac ar\Bigr).
\ee
Again, the parameter $a$ sets a characteristic length scale but is not associated with any physical size of the charge distribution. The electric field reads as
\be
E_{\rm log}(r)= {\frac{Q}{4\pi\varepsilon_0 }}\frac{1}{r(r+a)},
\ee
it obviously fits the Coulomb form at large distances, and the associated charge density
reads as
\be
\rho_{\rm log}(r)=\frac{Q}{4\pi}\frac{a}{r^2(r+a)^2}.
\ee
Its integral over the whole space sums to
\be 
4\pi\int_0^\infty r^2\rho_{\rm log}(r)\,dr=Q,
\ee
giving sense to the choice of the constant $\phi_0$ above. 
\textout{Assuming that the energy density could scale as in the theory of Maxwell, with $u_{\rm es}=\frac 12\varepsilon_0 E_{\rm log}^2(r)$, leads to a finite result once integrated over the whole space
\be
U_{\rm es}=\frac{Q^2}{8\pi\varepsilon_0}.
\ee}

This is therefore a very interesting model, recovering the Coulomb form at large distances but with a weaker divergence near the origin, presumably leading to a finite self-energy. The last question still deserves attention, since we have not yet written the Lagrangian density associated with this logarithmic model. Consequently, we do not yet know how to write the energy density $u_{\mathrm{log}}^{\mathrm{e}s}$ because we lack the function \(V_{\rm log}(\phi(r))\) entering
 \be{\mathscr L}_{\rm NL}=\frac 12\varepsilon_0 \Bigl(\frac {d\phi}{dr}\Bigr)^2-V_{\rm log}(\phi(r)),\ee   
 such that the equation of motion   
 \be \frac{d^2\phi}{dr^2}+\frac 2r\frac {d\phi}{dr}+\frac{1}{\varepsilon_0}\frac{dV_{\rm log}}{d\phi }=0, \ee
admits $\phi_{\rm log}(r)=\phi_0\ln\Bigl(1+\frac ar\Bigr)$ as a solution.
In order to find the correct $V_{\rm log}(\phi(r))$, if any, we form, from the solution
$\phi_{\rm log}(r)$ the quantity
\be
\frac{d^2\phi}{dr^2}+\frac 2r\frac {d\phi}{dr}=-\frac{\phi_0a^2}{r^2(r+a)^2},
 \ee
 and look for a solution of 
 \be
 \frac{1}{\varepsilon_0}\frac{dV_{\rm log}}{d\phi}=\frac{\phi_0a^2}{r^2(r+a)^2}
.
 \ee
 This is achieved by inverting 
 \be r=\frac{a}{e^{\phi/\phi_0}-1},\ee
 leading to
 \be
 \frac{dV_{\rm log}}{d\phi}=\frac{16 \phi_0\varepsilon_0}{a^2}\sinh^4\Bigl(\frac{\phi}{2\phi_0}\Bigr),
 \ee
 the solution of which is found to be 
 \be
 V_{\rm log}(\phi(r))=\frac{\phi_0^2\varepsilon_0}{a^2}\Bigl[
 \sinh\Bigl(\frac{2\phi}{\phi_0}\Bigr)-8 \sinh\Bigl(\frac{\phi}{\phi_0}\Bigr)+6\Bigl(\frac{\phi}{\phi_0} \Bigr)
 \Bigr]+V_0.
 \ee  
 This solves the problem and leads to a closed form to the effective charge density associated with the logarithmic potential
 \be 
 \rho_{\rm eff}=\frac{4\varepsilon_0\phi_0}{a^2}\Bigl( 
 \cosh\frac{\phi}{\phi_0}-1
 \Bigr)^2.
 \ee 
Using Derrick's theorem and Eq.~\eqref{Eq-Energy1}, we have 
\bea
U_{\rm NL}^{\rm es}[\phi]&=&\int d^3r\,\Bigl(
\frac{\phi_0^2\varepsilon_0}{a^2}\Bigl[
 \sinh\Bigl(\frac{2\phi}{\phi_0}\Bigr)-8 \sinh\Bigl(\frac{\phi}{\phi_0}\Bigr)+6\Bigl(\frac{\phi}{\phi_0} \Bigr)
 \Bigr] -\frac12\varepsilon_0\Bigl(\frac{d\phi}{dr}\Bigr)^2
\Bigr).
\eea 
 Eq.~\eqref{Eq-Energy3} is still shorter
and we find
\bea 
&&  U_{\rm log}^{\rm es}= - \frac{Q^2}{12\pi\varepsilon_0}\underbrace{\int_0^\infty \frac{dr}{(r+a)^2}}_{\frac 1a}
=-\frac{Q^2}{12\pi\varepsilon_0 a}.\eea 
The result is finite, but {\it negative}.

\subsection{Graphical representations}

\begin{figure}[!h]
 \centering
        \includegraphics[width=0.6\textwidth]{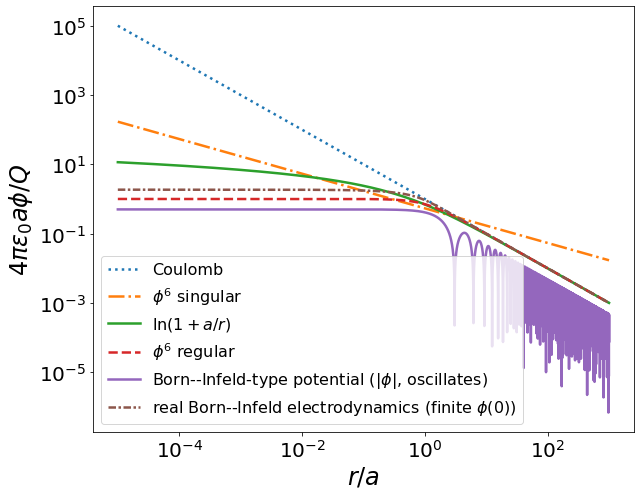}
 \caption{Electrostatic scalar potential $\phi(r)$ for nonlinear Mie electrodynamics compared to the Coulomb case (dotted line). From top to bottom, at small values of $r/a$, we plot 1) the Coulomb potential (dotted), 2) the singular solution associated with the $\phi^6$ potential (dot-dashed), 3) the logarithmic solution (solid), 
 4) the ``real'' Born--Infeld theory, studied in Sec.~\ref{sec4} (dot-dashed),
 5) the regular solution associated with the $\phi^6$ potential  (dashed), and 6) the Born--Infeld type of solution  [see SM] associated with the square-root potential  (solid). In this latter case, the solution is obtained numerically.}
\label{fig1}
\end{figure}

\begin{figure}[!h]
 \centering
        \includegraphics[width=0.6\textwidth]{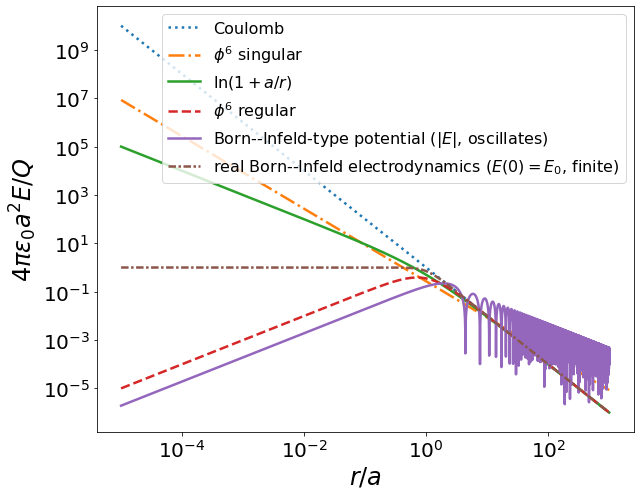}
 \caption{Electrostatic field $E(r)$ for nonlinear Mie electrodynamics compared to the Coulomb case (dotted line). From top to bottom at small values of $r/a$, we plot 1. the Coulomb {field} (dotted), 2. the singular solution associated with the $\phi^6$ potential  (dot-dashed), 3. the logarithmic solution (solid), 4. the ``real'' Born--Infeld theory, studied in Sec.~\ref{sec4} (dot-dashed), 5. the regular solution associated with the $\phi^6$ potential  (dashed), and 6. the Born--Infeld type of solution associated with the square-root potential  (solid). In this latter case, we obtain the solution numerically.}
\label{fig2}
\end{figure}

\begin{figure}[!h]
 \centering
        \includegraphics[width=0.6\textwidth]{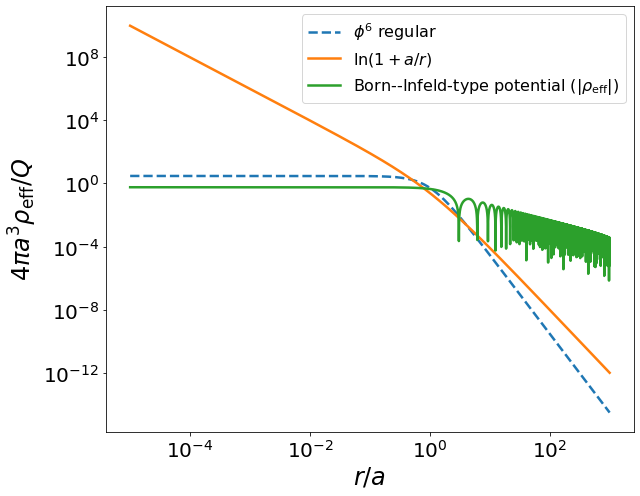}
 \caption{Effective charge densities $\rho_{\rm eff}(r)$ (from top to bottom at small values of $r/a$) 1. the logarithmic solution (solid), 2. the regular solution associated with the $\phi^6$ potential  (dashed), and 3. the Born--Infeld type of solution associated with the square-root potential  (solid). In the latter case, we obtain the solution numerically.}
\label{fig3}
\end{figure}

\begin{figure}[!h]
 \centering
        \includegraphics[width=0.6\textwidth]{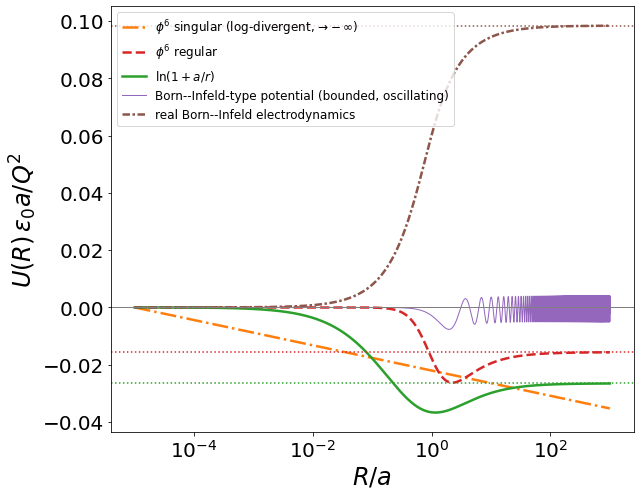}
 \caption{Total energy in a spherical shell of extension $R$,  for (from top to bottom at large values of $R/a$) 1. the ``real'' Born--Infeld theory, studied in Sec.~\ref{sec4} (dot-dashed), 2. the Born--Infeld type of solution  [see SM] associated with the square-root potential function (solid, solved numerically), 3. the regular solution associated with the $\phi^6$ potential function (dashed), 4.  the logarithmic solution (solid), 5. the singular solution associated with the $\phi^6$ potential function (dot-dashed).}
\label{fig4}
\end{figure}

Figs.~\ref{fig1}-\ref{fig4} illustrate the various cases considered. In particular, we note the oscillations of the ``Born--Infeld type'' potential in the large-distance--small-potential region [see SM].

In Fig.~\ref{fig1}, we plot the scaled electrostatic potential $\phi(r)$ for the various cases considered in section~\ref{sec3}. We superimpose the Born--Infeld solution discussed in Sec.~\ref{sec4}.

Fig.~\ref{fig2} shows the associated electric field for all cases, and Fig.~\ref {fig3} shows the effective charge densities for the three main problems.

Finally, the total energies in a spherical shell of extension $R$, defined by
\be U_{\rm NL}^{\rm es}=4\pi\int_0^R u_{\rm NL}^{\rm es}
r^2dr,\ee 
are plotted.
{For the regular \(\phi^6\) and logarithmic configurations, for which the surface term vanishes, the Mie construction yields a finite but negative total energy. The square-root candidate instead has oscillatory asymptotics (see SM), and its accumulated energy, while finite and positive, does not approach a well-defined limiting value.}

\subsection{Crucial flaws: the loss of gauge invariance and the negative total energy}

 It is worth noting here that Mie's theory predates Noether's theorem; consequently, the loss of gauge symmetry was not considered then as critical as it is today, since the link between gauge symmetry and electric charge conservation had not yet been elucidated, and the role of this symmetry in constructing Lagrangians for gauge interactions remained largely unexplored at the time.

Let us quote again
 Iwo Bia{\l}ynicki-Birula, who excludes Mie's theory on the basis of the loss of 
gauge invariance:
\begin{quotation}
``{\em The generation of the current from the field variables had been achieved in this manner, but at a very high price: the potentials acquired a direct physical meaning, and the gauge invariance was totally lost. This has been found unacceptable, and the theory of Mie was shelved for two decades.}''
\end{quotation} 

The situation with respect to the breaking of gauge invariance is very similar to the later theory of Proca\cite{Proca:1936_MassiveField}
and one might question the value of a non-gauge-invariant electrodynamic theory, particularly one developed after Weyl's work\cite{Weyl:1929_Elektron}.
Nevertheless, this critique is a commonly cited shortcoming of Proca's theory.  Indeed, the literature rarely notes that Proca intended to construct a relativistic theory based not on spinors, as in Dirac's theory \cite{dirac1928quantum1}, but on 4-vectors describing matter fields. In this context, the curl of this 4-vector field naturally emerges as an antisymmetric rank-2 tensor, making the analogy with the Faraday tensor immediate; however, it bears repeating that treating the electromagnetic field was not Proca's original intention.

Shortly after Mie published his trilogy of papers utilizing the Hamiltonian formalism, Born~\cite{Born1914} carefully analyzed the theory and reformulated it within the Lagrangian framework (see Ref.~\cite{SmeenkMartin2007} for a historical perspective). In particular, Born defined the total energy from the energy-momentum tensor, which ultimately yields a fatal negative value. This seems to be a robust result but, to our knowledge, does not appear  clearly stated in the literature. This raises serious questions, since the total energy should be identified with $mc^2$, where $m$ is the expected particle mass. From this perspective, Proca theory might seem surprising, since its total energy remains positive. This divergence with Mie's theory arises because, in our notation, the potential can be written as \(V_{\rm Proca}(\phi) = \rho\phi - \frac{1}{2}\epsilon_0 m^2\phi^2\), which violates the convexity condition. However, this does not pose an issue, since the charge is introduced externally into the Lagrangian, just as in Maxwell's electrodynamics.

\section{Born--Infeld theory}\label{sec4}
The two decades referred to in the previous quotation from Bia{\l}ynicki-Birula span the period before Born\cite{Born:1933_ModifiedField,Born:1934_QuantumTheory}, when Born and Infeld\cite{BornInfeld:1933_Nature,BornInfeld:1934_Foundations} introduced a new nonlinear theory that did not suffer from the loss of gauge invariance.
This theory also addresses the divergence of the field at a point charge.  It was explicitly modeled on how special relativity caps velocities at $c$: Born wanted an analogous ``relativity of the field strength." The Lagrangian of a relativistic free particle indeed reads as
\be 
L=-mc^2\sqrt{1-\frac{|\vac v|^2}{c^2}},
\quad S[\vac r]=-mc^2\int dt\, \sqrt{1-\frac{|\vac v|^2}{c^2}},
\ee 
so if we consider the electrostatic regime in which one assumes the existence of a maximum electric field intensity $E_0$, a natural candidate for a Lagrangian density is therefore
\be 
{\mathscr L}=- \varepsilon_0 E_0^2\sqrt{1-\frac{|\vac E|^2}{E_0^2}}.
\ee 
The nonlinearity does not appear in the equation of motion from the potential term, but in the kinetic term $K(E^2)$ in \eqref{Eq-KEPE}, hence it is built from a gauge invariant quantity from the beginning. In a sense, this is the opposite of Mie's strategy, which kept the kinetic energy intact but modified the potential.

Incorporating the magnetic field and adding an external electric source, unlike in the theory of Mie, the Lagrangian density thus becomes 
\be
{\mathscr L}_{\rm BI}=\varepsilon_0 E_0^2\left[
1-\left(
1-\frac{|\vac E|^2-c^2|\vac B|^2}{E_0^2}
\right)^{1/2}
\right]-\rho\phi+\vac J\cdot\vac A. \label{Eq_LagBIEB}
\ee
The first $1$ inside the square bracket yields the appropriate Maxwell limit when one expands the square root for $E_0\to\infty$, and it restores a Lagrangian equal to zero when the fields $\vac E$ and $\vac B$ vanish. 
The  covariant form associated reads as
\be
{\mathscr L}_{\rm BI}=\varepsilon_0 E_0^2\left[ 1-
\left(1-\tfrac{2}{\varepsilon_0 E_0^2}\Bigl( -\tfrac{1}{4\mu_0}F\dd\alpha\beta F\uu\alpha\beta\Bigr)\right)^{1/2}
\right] -A_\alpha J^\alpha. \label{Eq_LagBIEBcov}
\ee 

Note that this form is not the most general; Born and Infeld also considered additional terms associated with contracting the Faraday tensor with its dual pseudo-tensor, which we do not discuss here in the electrostatic regime\cite{Pauli:1981_Relativity}.

It is instructive to quote Born and Infeld to understand their motivations:
\begin{quotation}\begingroup 
{\em ``In the papers cited above, the new field theory has been introduced rather enigmatically. (\dots) The failure in the present theory may be expressed by the statement that it violates the principle of {\em finiteness} which postulates that a satisfactory theory should avoid letting physical quantities become infinite. Applying this principle to velocity leads to assuming an upper limit $c$ and replacing the Newtonian action function $\tfrac 12 mv^2$ of a free particle with the relativistic expression $mc^2(1-\sqrt{1-v^2/c^2})$. (\dots) Applying it to the electromagnetic field, one is led immediately to the assumption of an upper limit of the field strength and to a modification of the action function into
\[
{\mathscr L}=b^2\Bigl(  \sqrt{1+\tfrac{1}{b^2}(|\vac H|^2-|\vac E|^2)} -1 \Bigr).
\]
(\dots) Therefore, the new field theory seems to be a consequence of this very general principle, and the old one not more than a practical approximation, just in the same way as for the mechanics of Newton and Einstein.''   }
   \endgroup\end{quotation}

We also have, as usual, 
\be\vac E=-\grad\phi-\partial_t\vac A,\quad \vac B=\rot\vac A.\ee
Therefore, the first pair of Maxwell's equations remains unchanged,
\be \rot\vac E+\partial_t\vac B=0, \quad  \divergence\vac B=0.\ee
For the equations in the presence of sources, we can calculate from Eq.~\eqref{Eq_LagBIEB} 
\bea
&&\vac D=\frac{\partial \mathscr L_{\rm BI}}{\partial\vac E}=\varepsilon_0\vac E \left(
1-\frac{|\vac E|^2-c^2|\vac B|^2}{E_0^2}
\right)^{-1/2},\\
&&\vac H=-\frac{\partial \mathscr L_{\rm BI}}{\partial\vac B}=\frac{\vac B}{\mu_0} \left(
1-\frac{|\vac E|^2-c^2|\vac B|^2}{E_0^2}
\right)^{-1/2}.
\eea
In terms of these auxiliary fields, the equations of motion keep their usual form, 
\be \divergence\vac D =\rho, \quad\hbox{and}\quad \rot\vac H-\partial_t\vac D=\vac J.\ee
but in terms of the field strengths
they are clearly non-linear.
We have already argued that this generalization of Maxwell's equations has been introduced to prevent infinities in the field strengths due to point-like sources, and we treat a simple electrostatic example below. 
The electrostatic Lagrangian density is expressed in terms of the scalar potential as:
    \be
{\mathscr L}_{\rm BI}(\phi,\grad\phi)={-\varepsilon_0 E_0^2\left[1-\Bigl(
1-\frac{|\grad\phi(\vac r)|^2}{E_0^2}
\Bigr)^{1/2}\right]}-\rho\phi(\vac r),
\ee
leading to a modified Poisson equation:
    \be
\divergence\Bigl(
\grad\phi(\vac r) \Bigl(
1-\frac{|\grad\phi(\vac r)|^2}{E_0^2}
\Bigr)^{-1/2}
\Bigr)=-\rho(\vac r)/\varepsilon_0.
\ee

In the case of a point charge $q$ at the origin, this 
takes the form, for $r\not=0$,
\be
\frac{1}{r^2}\frac d{dr}\Bigl[
r^2\frac{d\phi}{dr}\Bigl(
1-\frac 1{E_0^2}\Bigl(\frac{d\phi}{dr}\Bigr)^2
\Bigr)^{-1/2}
\Bigr]=0,
\ee
but we rather use Gauss law in the form $\divergence\vac D=Q\delta^{(3)}(\vac r)$ to write
\be \vac D(\vac r)=\frac{Q}{4\pi r^2}\hat{\vac r}\ee
from which we deduce
\be
\vac E(\vac r)=\frac{E_0}{\sqrt{1+(r/r_0)^4}} \hat{\vac r}\label{EqEBI}
\ee 
with {$r_0^2=|Q|/(4\pi\varepsilon_0 E_0)$}, $r_0$ is the Born--Infeld radius.

The resulting electric field
\be
E(r)=\frac{Q}{4\pi\varepsilon_0r^2}\left(
1+\frac{Q^2}{16\pi^2\varepsilon_0^2E_0^2r^4}
\right)^{-1/2}
\ee
is Coulombic at large distances and
has, at short distances, the form
\be
E(r)\simeq E_0(1-({8\pi^2\varepsilon_0^2E_0^2}/{Q^2})r^4),\quad r\to 0
\ee
bounded as expected by the constant $E_0$. This case is simple to solve, but the modified Poisson equation --- and the associated modified Maxwell equation --- are clearly nonlinear and difficult to solve in general.

The main difficulty of Born and Infeld's theory is indeed its nonlinear character in the $\vac E$ and $\vac B$ fields. This makes it much harder to solve problems and the principle of superposition breaks down; despite Born and Infeld's enthusiasm, this is a very high price to pay!
Nevertheless, this theory addresses the question of the self-energy correctly. Indeed, following the procedure described earlier for the Maxwell and Mie theories, we can build 
the stress-energy tensor from Eq.~\eqref{Eq_LagBIEBcov}. 
The main difference comes from the term
\be 
K(Q_2)= \varepsilon_0 E_0^2\left[ 1-
\left(1-\tfrac{2}{\varepsilon_0 E_0^2}\Bigl( -\tfrac{1}{4\mu_0}F\dd\alpha\beta F\uu\alpha\beta\Bigr)\right)^{1/2}
\right],
\ee 
from which
\be
2\frac{\partial K}{\partial g\uu\alpha\beta}=K'(Q_2)\tfrac{-1}{\mu_0}F\dd\alpha\rho F\du\beta\rho,
\ee
{taking the $00$ component we get
\be
2\frac{\partial K}{\partial g\uu00}=\frac{1 }{\sqrt{1-\frac{|\vac E|^2-c^2|\vac B|^2}{E_0^2}}}
\tfrac{1}{\mu_0}(|\vac E|^2/c^2-|\vac B|^2).
\ee }
Collecting the terms and the electrostatic limit of the $00$ component of the stress tensor leads to the compact form
\be 
T\dd 00^{\rm BI}=\varepsilon_0 E_0^2
\left[
\frac{1 }{\sqrt{1-\frac{|\vac E|^2 }{E_0^2}}}-1\right].
\ee

Using the form obtained in Eq.~\eqref{EqEBI}, we write
\bea 
U_{\rm BI}^{\rm es}[\phi]&=&4\pi \varepsilon_0 E_0^2\int_0^\infty dr\, r^2\Bigl(
\sqrt{1+(r_0/r)^4}-1
\Bigr),\nonumber\\ &=& 
\frac{2\sqrt\pi}{3}\Gamma^2(1/4) \varepsilon_0 E_0^2 r_0^3
=
{\frac{\Gamma^2(1/4)}{6\sqrt\pi}|Q|E_0r_0},
\eea 
where we have used 
$\int_0^\infty dx\, x^2 (\sqrt{1+1/x^4}-1)=\Gamma^2(1/4)/(6\sqrt \pi)\simeq 1.236049\dots$.
The electrostatic self-energy of a point charge is now finite, as Born and Infeld expected.
  This scales as in the classical finite-sphere model. Setting $U_{\rm BI}^{\rm es}=mc^2$
 fixes $r_0\sim Q^2/(\varepsilon_0mc^2)$, the same order of magnitude as the classical electron radius (if $Q=-|q_e|$), up to the numerical factor.

Figs.~\ref {fig1} and \ref{fig2} show the electrostatic potential and field of Born--Infeld's theory compared with those of various Mie theories.

\section{Discussion}

This is not the end of the story with nonlinear extensions of electrodynamics.
In 1936,  Heisenberg and Euler\cite{HeisenbergEuler:1936} 
considered quantum corrections to electrodynamics and the role of the virtual electron-positron pairs.

For fields much weaker than the critical value  $E_c = \frac{m^2c^3}{e\hbar} \approx 1.3 \times 10^{18}\ \text{V/m}$, a perturbative expansion
 produces non-linearities in the theory. The effective Lagrangian  density (here we fix {$\hbar=\varepsilon_0=c=1$)} reads as\cite{Zangwill2013}
\be
\mathscr L_{\rm HE}=\frac 12(|\vac E|^2-|\vac B|^2)+\frac{2\alpha^2}{45 m^4}[(|\vac E|^2-|\vac B|^2)^2+7(\vac E\cdot\vac B)^2]-\rho\phi+\vac J\cdot\vac A.
\ee
If we limit our considerations again to a static electric problem, the Euler--Lagrange equation simplifies into the form
\be
\frac{\partial \mathscr L_{\rm HE}}{\partial\phi}
-\divergence \Bigl(\frac{\partial \mathscr L_{\rm HE}}{\partial(\grad\phi)}\Bigr)=0
\ee
  with,  $\phi$ deduced from $\vac E=-\grad\phi$ and
$ \vac D=\frac{\partial \mathscr L_{\rm HE}}{\partial(-\grad\phi)}
$
to recover the standard Gauss equation
\be -\rho+\divergence\vac D=0,\ee 
except that now $\vac D$ deviates from its usual value $\vac D=\vac E$ in Maxwell's theory and is equal to
\be 
\vac D=\vac E+\kappa [(|\vac E|^2-|\vac B|^2)\vac E
+(7/2)(\vac E\cdot\vac B)\vac B],
\ee 
with {$\kappa=8\alpha^2/(45m^4)$.}

For the simple case of a purely electrostatic problem of the point charge located at the origin, with $\vac E=E(r)\hat{\vac r}$ and $\vac B=0$, the Gauss theorem works in terms of $\vac D=D(r)\hat{\vac r}$ in the standard manner 
\bea 
&&\frac 1{r^2}\frac d{dr}\bigl( r^2 D(r)\bigr)={q}\delta^{(3)}(\vac r),\\
&& D(r)=\frac q{4\pi r^2}.
\eea 
Now, we use 
\be 
D(r)=E(r)+\kappa E^3(r)
\ee 
and write $E(r)$ as
 $E(r)={E_C}(r)+\delta E(r)$, ${E_C}(r)=q/(4\pi r^2)$ being the standard Coulomb result. The expression above for $D(r)$ leads, in the perturbative limit $3\kappa {E_C}^2(r)\ll 1$, to
 \be 
 \delta E(r)\simeq -\kappa {E_C}(r)^3=-\kappa\Bigl(\frac q{4\pi r^2}\Bigr)^3,
 \ee 
i.e. a $r^{-6}$ attenuation to Coulomb law, {signaling the breakdown of the weak-field Euler–Heisenberg expansion near the point charge. It therefore cannot be used to determine the short-distance self-energy. In full QED the electron self-energy is treated through renormalization.}

The key takeaway for students is likely that the electrodynamics theory they encountered in their studies, Maxwell's theory, is not the only formal possibility for describing electromagnetic phenomena. Moreover, it suffers from certain inconsistencies that specific nonlinear extensions may address.
Perhaps even more importantly, nonlinear corrections are required to account for certain quantum effects. This serves as an instructive example of how physics evolves, overcoming past shortcomings through successive theoretical refinements.
  
\bibliographystyle{iopart-num}
\bibliography{references}

\providecommand{\newblock}{}
\begin{thebibliography}{10}
\expandafter\ifx\csname url\endcsname\relax
  \def\url#1{{\tt #1}}\fi
\expandafter\ifx\csname urlprefix\endcsname\relax\def\urlprefix{URL }\fi
\providecommand{\eprint}[2][]{\url{#2}}

\bibitem{maxwell1873treatise}
Maxwell J~C 1873 {\em A Treatise on Electricity and Magnetism\/} 1st ed vol 1-2
  (Oxford: Clarendon Press) vol. 1 online at
  \url{https://archive.org/details/electricandmagne01maxwrich}, Vol. 2 online
  at \url{https://www.aproged.pt/biblioteca/MaxwellII.pdf}

\bibitem{Thomson1881}
Thomson J~J 1881 {\em Philosophical Magazine Series 5\/} {\bf 11} 229--249
  \urlprefix\url{https://www.tandfonline.com/doi/abs/10.1080/14786448108627008}

\bibitem{Landau1980Classical}
Landau L~D and Lifshitz E~M 1980 {\em The Classical Theory of Fields\/} 4th ed
  ({\em Course of Theoretical Physics\/} vol~2) (Oxford: Butterworth-Heinemann)
  ISBN 978-0750627689

\bibitem{Berche2025-10}
Berche B and Medina E 2025 {\em Alternative Relativistic Field Theories\/}
  (Cham: Springer Nature Switzerland) pp 427--495 ISBN 978-3-031-78962-5
  \urlprefix\url{https://doi.org/10.1007/978-3-031-78962-5_10}

\bibitem{Franklin2010}
Franklin J 2010 {\em Advanced Mechanics and General Relativity\/} (Cambridge:
  Cambridge University Press) ISBN 978-0-521-76245-8

\bibitem{Berche2025-4}
Berche B and Medina E 2025 {\em Spacetime Symmetries in the Minkowski
  Manifold\/} (Cham: Springer Nature Switzerland) pp 117--151 ISBN
  978-3-031-78962-5 \urlprefix\url{https://doi.org/10.1007/978-3-031-78962-5_4}

\bibitem{Zangwill2013}
Zangwill A 2013 {\em Modern Electrodynamics\/} (Cambridge: Cambridge University
  Press) ISBN 978-0-521-89697-9

\bibitem{Tong2025Electromagnetism}
Tong D 2025 {\em Electromagnetism\/} ({\em Lectures on Theoretical Physics\/}
  vol~2) (Cambridge: Cambridge University Press) ISBN 978-1-009-59462-2

\bibitem{Berche2025-5}
Berche B and Medina E 2025 {\em Relativistic Field Theory\/} (Cham: Springer
  Nature Switzerland) pp 153--214 ISBN 978-3-031-78962-5
  \urlprefix\url{https://doi.org/10.1007/978-3-031-78962-5_5}

\bibitem{Schwarzschild1903a}
Schwarzschild K 1903 {\em Nachrichten von der Gesellschaft der Wissenschaften
  zu G{\"o}ttingen, Mathematisch-Physikalische Klasse\/} {\bf 1903} 126--131
  \urlprefix\url{http://eudml.org/doc/58546}

\bibitem{Schwarzschild1903b}
Schwarzschild K 1903 {\em Nachrichten von der Gesellschaft der Wissenschaften
  zu G{\"o}ttingen, Mathematisch-Physikalische Klasse\/} {\bf 1903} 132--141
  \urlprefix\url{http://eudml.org/doc/58547}

\bibitem{10.1119/1.4955153}
Berche B, Malterre D and Medina E 2016 {\em American Journal of Physics\/} {\bf
  84} 616--625 ISSN 0002-9505 \urlprefix\url{https://doi.org/10.1119/1.4955153}

\bibitem{griffiths_electrodynamics_2024}
Griffiths D~J 2024 {\em Introduction to Electrodynamics\/} 5th ed (Cambridge
  University Press) ISBN 9781009397759

\bibitem{Mie:1912_Part1}
Mie G 1912 {\em Annalen der Physik\/} {\bf 342} 511--534
  \urlprefix\url{https://doi.org/10.1002/andp.19123420306}

\bibitem{Mie:1912_Part2}
Mie G 1912 {\em Annalen der Physik\/} {\bf 344} 1--40
  \urlprefix\url{https://doi.org/10.1002/andp.19123441102}

\bibitem{Mie:1913_Part3}
Mie G 1913 {\em Annalen der Physik\/} {\bf 345} 1--66
  \urlprefix\url{https://doi.org/10.1002/andp.19133450102}

\bibitem{Pauli:1981_Relativity}
Pauli W 1981 {\em Theory of Relativity\/} (New York: Dover Publications) ISBN
  978-0-486-64152-2 reprint of the 1958 English translation published by
  Pergamon Press; originally published in German in 1921

\bibitem{Vizgin1994}
Vizgin V~P 1994 {\em Unified Field Theories in the First Third of the 20th
  Century\/} ({\em Science Networks. Historical Studies\/} vol~13) (Birkhäuser
  Basel) ISBN 978-3-7643-2679-1

\bibitem{BialynickiBirula1983}
Bia{\l}ynicki-Birula I 1983 Nonlinear electrodynamics: Variations on a theme by
  {B}orn and {I}nfeld {\em Quantum Theory of Particles and Fields: Birthday
  Volume Dedicated to Jan {\L}opusza{\'n}ski\/} ed Jancewicz B and Lukierski J
  (Singapore: World Scientific) pp 31--48 ISBN 978-9971-950-77-4

\bibitem{Proca:1936_MassiveField}
Proca A 1936 {\em Journal de Physique et le Radium\/} {\bf 7} 347--353
  \urlprefix\url{https://doi.org/10.1051/jphysrad:0193600708034700}

\bibitem{wald2022advanced}
Wald R~M 2022 {\em Advanced Classical Electromagnetism\/} (Princeton, NJ:
  Princeton University Press) ISBN 978-0-691-22039-0

\bibitem{Berche2025-9}
Berche B and Medina E 2025 {\em Curved Spacetime Field Theory\/} (Cham:
  Springer Nature Switzerland) pp 331--425 ISBN 978-3-031-78962-5
  \urlprefix\url{https://doi.org/10.1007/978-3-031-78962-5_9}

\bibitem{pohozaev1965eigenfunctions}
Pohozaev S~I 1965 {\em Soviet Math. Dokl.\/} {\bf 6} 1408--1411

\bibitem{Derrick:1964_NonlinearWave}
Derrick G~H 1964 {\em Journal of Mathematical Physics\/} {\bf 5} 1252--1254
  \urlprefix\url{https://doi.org/10.1063/1.1704233}

\bibitem{Mach_2012}
Mach P 2012 {\em Journal of Mathematical Physics\/} {\bf 53} ISSN 1089-7658
  \urlprefix\url{http://dx.doi.org/10.1063/1.4725414}

\bibitem{Weyl:1929_Elektron}
Weyl H 1929 {\em Zeitschrift f{\"u}r Physik\/} {\bf 56} 330--352
  \urlprefix\url{https://doi.org/10.1007/BF01339504}

\bibitem{dirac1928quantum1}
Dirac P~A~M 1928 {\em Proceedings of the Royal Society of London. Series A,
  Containing Papers of a Mathematical and Physical Character\/} {\bf 117}
  610--624

\bibitem{Born1914}
Born M 1914 {\em Nachrichten von der Gesellschaft der Wissenschaften zu
  Göttingen, Mathematisch-Physikalische Klasse\/} {\bf 1914} 23--36
  \urlprefix\url{https://eudml.org/doc/58898}

\bibitem{SmeenkMartin2007}
Smeenk C and Martin C 2007 Mie's theories of matter and gravitation {\em The
  Genesis of General Relativity: Alternative Approaches to General
  Relativity\/} ({\em Boston Studies in the Philosophy of Science\/} vol 250)
  ed Renn J (Dordrecht: Springer) pp 1543--1553

\bibitem{Born:1933_ModifiedField}
Born M 1933 {\em Nature\/} {\bf 132} 282
  \urlprefix\url{https://doi.org/10.1038/132282a0}

\bibitem{Born:1934_QuantumTheory}
Born M 1934 {\em Proceedings of the Royal Society of London. Series A,
  Containing Papers of a Mathematical and Physical Character\/} {\bf 143}
  410--437 \urlprefix\url{https://doi.org/10.1098/rspa.1934.0010}

\bibitem{BornInfeld:1933_Nature}
Born M and Infeld L 1933 {\em Nature\/} {\bf 132} 1004
  \urlprefix\url{https://doi.org/10.1038/1321004b0}

\bibitem{BornInfeld:1934_Foundations}
Born M and Infeld L 1934 {\em Proceedings of the Royal Society of London.
  Series A, Containing Papers of a Mathematical and Physical Character\/} {\bf
  144} 425--451 \urlprefix\url{https://doi.org/10.1098/rspa.1934.0059}

\bibitem{HeisenbergEuler:1936}
Heisenberg W and Euler H 1936 {\em Zeitschrift f{\"u}r Physik\/} {\bf 98}
  714--732 \urlprefix\url{https://doi.org/10.1007/BF01343663}

\end{thebibliography}

\end{document}